\documentstyle[11pt,newpasp,twoside,epsf]{article}
\def\edcomment#1{\iffalse\marginpar{\raggedright\sl#1\/}\else\relax\fi}
\reversemarginpar
\begin{document}
\title{HI and Hot Gas in the Outskirts of the M81 Group}
\author{M.\ Bureau}
\affil{Hubble Fellow, Columbia Astrophysics Laboratory, 550 West 120th
 Street, 1027 Pupin Hall, Mail Code 5247, New York, NY~10027, U.S.A.}
\author{F.\ Walter}
\affil{Jansky Fellow, NRAO, P.O.\ Box~O, Socorro, NM~87801, U.S.A.}
\author{J.\ van Gorkom}
\affil{Department of Astronomy, Columbia University, 538 West 120th
 Street, New York, NY~10027, U.S.A.}
\author{C.\ Carignan}
\affil{D\'{e}partement de Physique, Universit\'{e} de Montr\'{e}al,
 C.P.\ 6128, Succ.\ Centre-Ville, Montr\'{e}al, QC H3C 3J7, Canada}
%
% Abstract
%
\begin{abstract}
Results are presented from a wide area, high resolution HI synthesis
survey of the outer regions of the nearby M81 group, where internal
(galactic) and external (group-related) evolution processes can be
studied simultaneously in great detail. The survey encompasses the
star forming dwarf galaxies M81dwA, UGC4483, and HoII, where evidence
of ram pressure stripping was recently discovered. The data do not
reveal any intergalactic HI, but the outer parts of HoII are
reminiscent of tidal tails. We argue however that those structures are
equally consistent with the latest ram pressure models including
cooling. The case for a hot intergalactic medium in this poor,
spiral-only group is thus still open. The survey also puts tight
constraints on possible counterparts to the local high velocity cloud
population in an external group, reaching a $3\sigma$ column density
of $10^{19}$~atom~cm$^{-2}$ and a $6\sigma$ limiting mass of
$1.5\times10^5$~$M_{\mbox{\scriptsize \sun}}$.
\end{abstract}
%
% Introduction
%
\section{Introduction}
\label{sec:intro}
Most galaxies in the local universe are located in groups, where
dwarfs dominate by number. Understanding the influence of this
environment on (dwarf) members is therefore essential to our
comprehension of galaxy formation and evolution. Nearby groups are
particularly attractive for this purpose since both internal
(galactic) and external (group-related) evolution processes can be
studied in details. Bureau \& Carignan (2002) presented neutral
hydrogen Very Large Array (VLA) D-array observations of the dwarf
irregular galaxy HoII, a prototype galaxy for studies of
self-propagating star formation (Puche et al.\ 1992). HI was detected
to radii over $16\arcmin$ or $4R_{25}$, with a total mass
$M_{\mbox{\tiny HI}}=6.44\times10^8$~$M_{\scriptsize \sun}$. Most
importantly, the integrated HI map revealed a characteristic
comet-like appearance, with a large but faint component extending to
the northwest and the HI appearing compressed on the opposite side,
suggesting that HoII is affected by ram pressure from an intragroup
medium (IGM).

HoII lies roughly $0.5$~Mpc northeast of the core of the nearby M81
Group ($D=3.2$~Mpc), along with the dwarf irregular galaxy M81dwA
(Kar52) and the blue compact dwarf galaxy UGC4483. No obvious signs of
interaction were detected, but the three galaxies are most likely part
of the NGC2403 subgroup, infalling toward M81 (Karachentsev et al.\
2000). With large uncertainties, ram pressure stripping of the outer
parts of the disk would require an IGM density $n_{\mbox{\tiny
IGM}}\ga4.0\times10^{-6}$~atoms~cm$^{-3}$ near HoII, or about $1$\% of
the virial mass of the group uniformly distributed over a volume just
enclosing it, consistent with the known X-ray properties of small
groups. The presence of an IGM in such a poor group, lacking any
early-type galaxy, would have important consequences for the evolution
of group members, and it could explain why many HI holes in HoII are
located in low surface density regions of the disk, where no star
formation is expected or observed (Rhode et al.\ 1999; Bureau \&
Carignan 2002).
%
% Observations and Survey
%
\section{Survey}
\label{sec:survey}
To test whether the HI morphology of HoII is indeed due to ram
pressure stripping, and thus to the existence of an extended hot IGM,
or whether tidal interactions offer a better explanation, we have
conducted an HI survey of the area encompassing HoII, M81dwA, and
UGC4483. Our setup is entirely driven by the need for excellent
surface brightness sensitivity, and constraints on the presence of any
intergalactic HI clouds are obtained for free.

A mosaic of $36$ contiguous VLA D-array pointings was obtained over
$5$ separate observing runs, each pointing totaling roughly $50$~min
of integration on source. The mosaic partially covers an area of about
$3\deg\times3\deg$, with a pointing spacing of $22\arcmin$, ensuring
proper sampling of the $32\arcmin$ primary beam. A $3.125$~MHz
bandwidth with $64$ channels was used for all observations
($10.3$~km~s$^{-1}$ per channel; online Hanning smoothing), covering
the entire velocity range of the three galaxies and the M81 and
NGC2403 groups, while simultaneously resolving the internal kinematics
of individual galaxies. Except at the edges, the rms noise per channel
achieved is approximately uniform at $1.0$~mJy~beam$^{-1}$,
corresponding to $3.0\times10^{18}$~atoms~cm$^{-2}$ or
$2.5\times10^4$~$M_{\scriptsize \sun}$ at the distance
adopted. Preliminary total HI and intensity-weighted mean velocity
fields of the entire mosaic are presented in Figure~1, while blow-ups
of the HoII--M81dwA pair are shown in Figures~2 and 3. The synthesized
beam is $65\arcsec\times65\arcsec$ in all maps.
%
% Large-Scale Structure and Intergalactic HI
%
\section{Large-Scale Structure and Intergalactic HI}
\label{sec:intergalactic}
The main result from the large-scale HI structure presented in
Figure~1 is that no HI is detected outside that directly associated
with HoII, M81dwA, and UGC4483. That is, there is no evidence of
either large-scale tidal interactions or any intergalactic HI cloud
population within a broad area encompassing the three galaxies (and
more).

\begin{figure}[t]
\plottwo{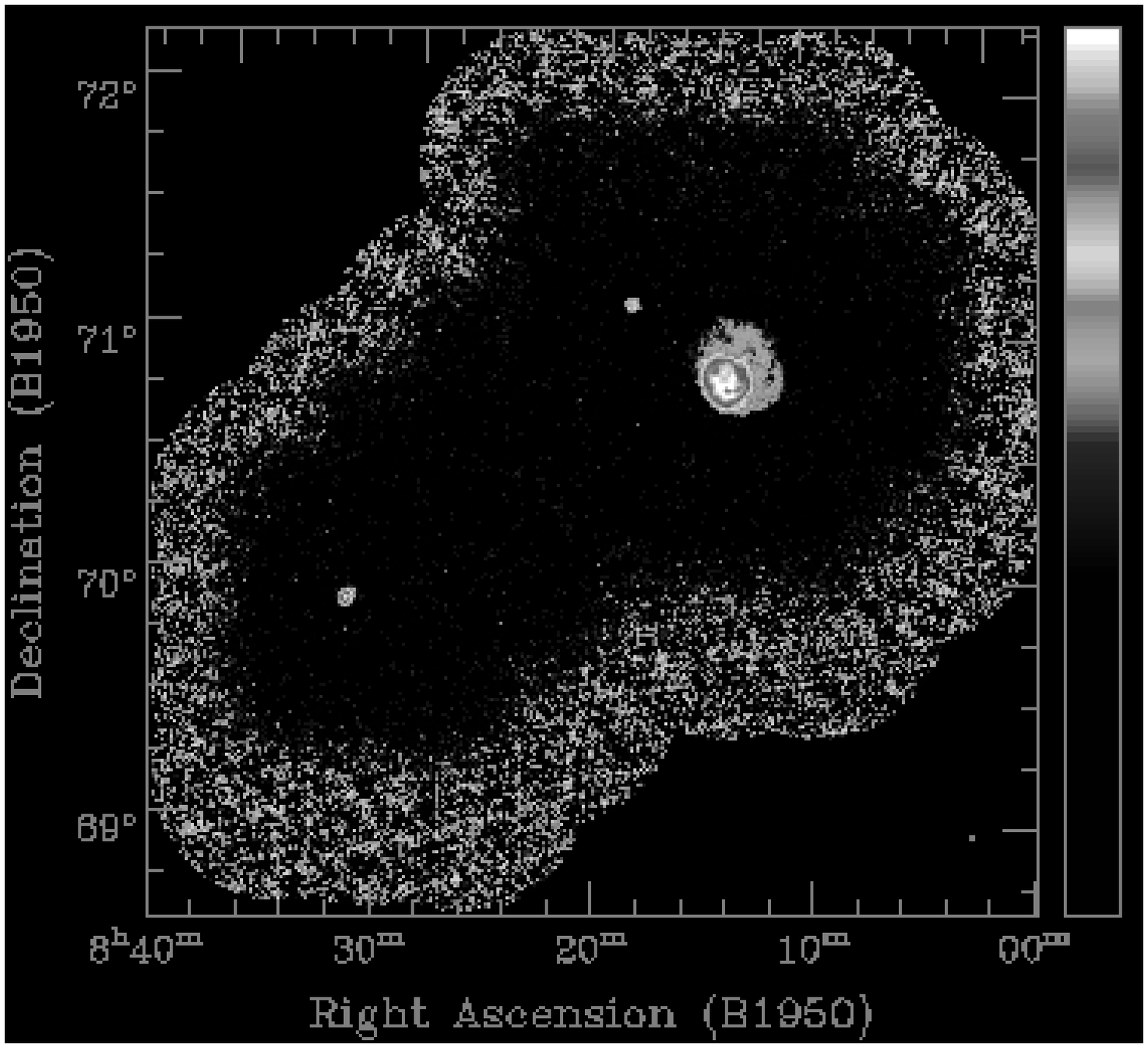}{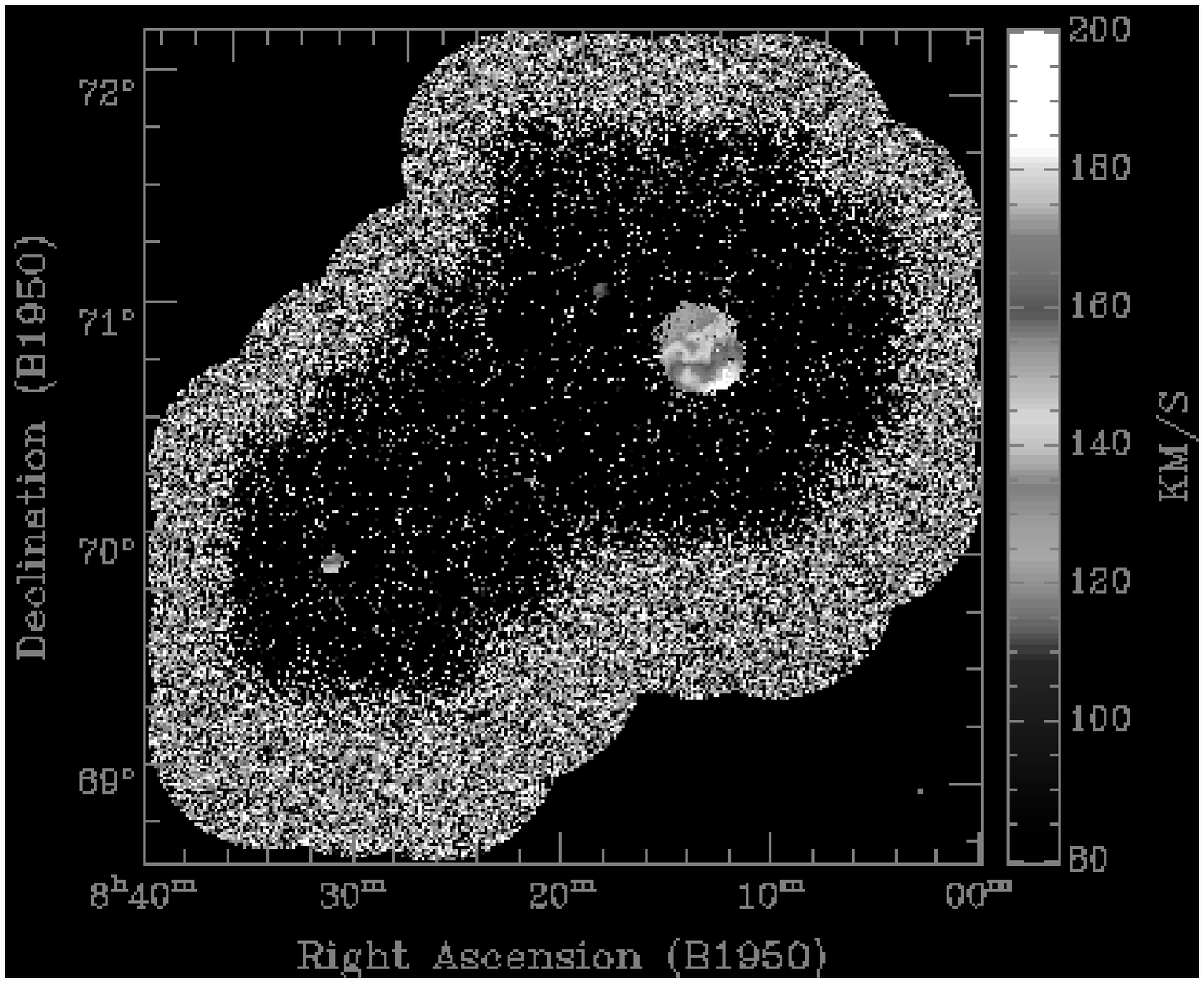}
\caption{{\em Left:} Total HI map of the entire survey area
(logarithmic fluxes). Note the absence of large-scale HI beside that
associated with the three galaxies. {\em Right:} Intensity-weighted
mean velocity field.}
\end{figure}
Our excellent surface brightness sensitivity ensures that we can
detect intergalactic HI down to column densities of about
$1.0\times10^{19}$~atoms~cm$^{-2}$ at the $3\sigma$ level, below which
the hydrogen is probably ionized by the metagalactic UV radiation
field. Across the entire mosaic, no HI is detected in any channel at a
level of $5\sigma$, corresponding to
$1.5\times10^{19}$~atoms~cm$^{-2}$ or $1.25\times10^5$~$M_{\scriptsize
\sun}$ (per beam, of $1$~kpc diameter at the distance
adopted). Although we have not yet carried out a proper statistical
analysis, no patch of material at a lower column density is detected
either, so tighter limits apply to extended material and limits on the
volume density of various intergalactic HI cloud populations will be
derived.

Those limits are extremely interesting considering the current debate
about the nature of the high velocity clouds (HVCs) surrounding the
Galaxy. Blitz et al.\ (1999) suggest that some HVCs are the long
sought minihalos overproduced in cold dark matter structure formation
simulations. A typical compact HVC has a mass of $4.5\,(\frac{D}{{\rm
kpc}})^2\times10^6$~$M_{\scriptsize \sun}$ and a size of
$0.4$~deg$^2$, corresponding to about $1\times10^6$~$M_{\scriptsize
\sun}$ and a diameter of $6$~kpc at the typical expected distance of
$500$~kpc ($N_{\mbox{\tiny HI}}=1\times10^{19}$~atoms~cm$^{-2}$ at
peak and $\sigma_{\mbox{v}}=25$~km~s$^{-1}$; Putman et al.\ 2002). If
present, counterparts to the Local Group HVCs should thus appear as
large patches of faint HI in our data (many beam widths). A cursory
examination of our cube reveals no such signal, so the analogs of
Local Group HVCs remain undetected in the M81 Group if located at
hundreds of kiloparsecs from the core (where our observations are
located and the limits become interesting). It remains to be seen
however if the low surface brightness material in the northwest half
of HoII would be seen as HVCs by an observer located within its disk.
%
% The HoII--M81dwA Pair
%
\section{HoII and M81dwA}
\label{sec:ho2}
Given the absence of large-scale HI or intergalactic clouds, we now
focus our attention on the HoII--M81dwA pair (Figs.~2--3). HoII itself
still shows a high surface brightness component which appears
axisymmetric (except for the presence of numerous holes and shells)
and is clearly rotationally supported. The sharp edge to the southwest
of the HI distribution is also very much present. The low surface
brightness component to the northwest, which except for its comet-like
morphology appeared structureless in the shallower data of Bureau \&
Carignan (2002), is now resolved into at least two and perhaps three
arms. In particular, the arm to the southwest can be traced through
over $\frac{1}{2}$ turns, and perhaps $1\frac{1}{2}$, in the channel
maps.

HoII's HI morphology is reminiscent of that expected from tidal
interactions, M81dwA providing a likely interacting companion. The arm
kinematics is however regular, M81dwA appears undisturbed, and no
bridge is visible between the two galaxies (or any debris outside of
their immediate vicinities). The case for a tidal interaction at the
origin of HoII's HI distribution and kinematics is thus not
bullet-proof, and the possibility of ram pressure stripping can not be
rejected outright. We explore each of those possibilities in more
details below.

\begin{figure}[t]
\plotone{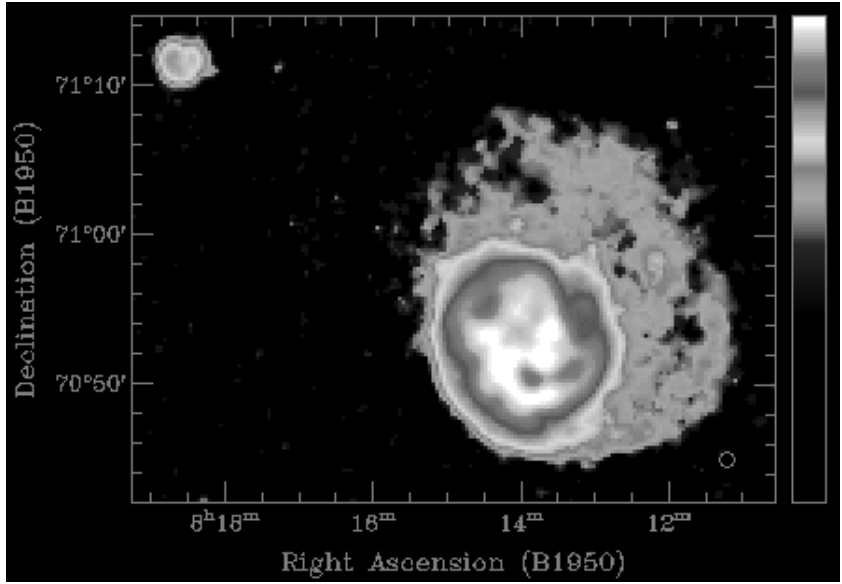}
\caption{Total HI map of the survey area near HoII and M81dwA
(logarithmic fluxes).}
\end{figure}
%
% Tidal Interactions
%
\subsection{Tidal Interactions}
\label{sec:interactions}
The HI distribution of HoII displays some of the characteristics
expected from the interaction of a disk galaxy with a passing lower
mass companion: strong asymmetry, sharp ``near'' tidal arm, and
broader ``far'' tidal arm. In such cases, the arms are material, as
appears to be the case here. The argument for an HoII--M81dwA tidal
interaction can be enlightened by a comparison with the well-studied
case of M51 and its companion NGC5195. In M51, the inner spiral arms
were probably caused by a close passage of NGC5195 (on a bound orbit),
while the outer tidal arm was most likely caused by a farther and
earlier passage (e.g.\ Howard \& Byrd 1990). The later in particular
is very similar to the morphology of HoII, being located well outside
of the optical body of the galaxy and composed essentially entirely of
HI (Rots et al.\ 1990). It is extended, one-sided, curved, with a
sharp edge on its leading side and clumps of HI at its end.

The HI clumps at the end of M51's tidal arm are also interesting in
that they are located roughly where M81dwA currently lies with respect
to HoII, and they are expected in a tidal model (Howard \& Byrd
1990). This raises the possibility that M81dwA is not the companion at
the origin of the disturbed morphology of HoII, but rather a tidal
dwarf, i.e.\ a low-mass galaxy formed through gravitational collapse
of the tidal debris from a previous interactions, perhaps with
UGC4483. This possibility is supported by the kinematics of M81dwA,
which appears to contain little if any dark matter (Sargent, Sancisi,
\& Lo 1983), but it must remain speculative until detailed modeling of
the HoII--M81dwA pair (and possibly of all three galaxies) is carried
out.

\begin{figure}[t]
\plotone{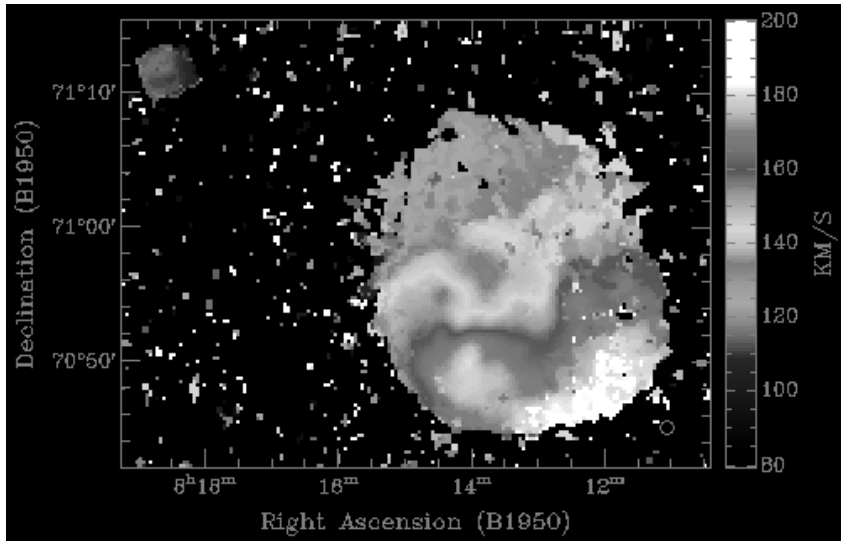}
\caption{Intensity-weighted mean HI velocity field of the survey area
near HoII and M81dwA.}
\end{figure}
%
% Ram Pressure Stripping
%
\subsection{Ram Pressure Stripping}
\label{sec:ram_pressure}
The arms detected within the northwest low surface brightness HI
component of HoII are not normally associated with ram pressure, which
tends to create clumpy and disorganized comet-like tails (e.g.\
Quilis, Moore, \& Bower 2000). However, recent models including gas
cooling do yield similar structures (Schulz \& Struck 2001). As usual,
a face-on galaxy--IGM encounter yields a symmetric gas distribution,
while an inclined encounter yields an asymmetric one. The latter is
indeed expected in the case of HoII given its location with respect to
the core of the M81 Group. Most importantly, when the compressed gas
(due to ram pressure) is allowed to cool, it becomes gravitationally
unstable, develops spiral arms, and looses angular momentum to the
outer disk, thus shrinking in the process. This process of
compression-driven annealing also leads to the formation of a sharp
edge and a dense annulus in the inner parts of the galaxy, in effect
shielding the center of the galaxy from further stripping, while
abandoning the outer disks to the IGM (Schulz \& Struck 2001). The
resulting morphology is very much like that observed in HoII (Fig.~2).

Given the coupling between the disk rotation and the IGM ``wind'' in
ram pressure simulations including cooling, it is not possible to
reject outright the possibility that ram pressure is indeed at the
origin of HoII's HI morphology.
%
% Conclusions
%
\section{Conclusions}
\label{sec:conclusions}
We have presented preliminary results from a wide-field but sensitive
HI synthesis survey of the outskirts of the M81 Group, in an area
encompassing the dwarf galaxies HoII, M81dwA, and UGC4483. No signs of
large-scale tidal features are present, and no intergalactic HI clouds
are detected, this to a level better than
$1.5\times10^{19}$~atoms~cm$^{-2}$ or $1.25\times10^5$~$M_{\scriptsize
\sun}$ (per channel and beam), which appears to rule out counterparts
to the Local Group HVCs if located at cosmologically relevant
distances. The HI distribution and kinematics of HoII are consistent
with that expected from tidal interactions, especially M51-like
events, but ram pressure models including gas cooling are also
possible. It thus appears premature to rule out either possibilities,
and thus the presence of a hot extended IGM in the M81 Group. It is
also possible that M81dwA is a tidal dwarf. A search for a (faint)
stellar counterpart to the northwest HI arms in HoII should resolve
the issue, given that ram pressure is only expected to affect the gas.
%
% Acknowledgments
%
\acknowledgments{Support for this work was provided by NASA through
Hubble Fellowship grant HST-HF-01136.01 awarded by the Space Telescope
Science Institute, which is operated by the Association of
Universities for Research in Astronomy Inc.\ for NASA under contract
NAS~5-26555, and by NSF grant AST~00-98249 to Columbia University. The
Digitized Sky Surveys were produced at the Space Telescope Science
Institute under U.S.\ Government grant NAG W-2166. The images of these
surveys are based on photographic data obtained using the Oschin
Schmidt Telescope on Palomar Mountain and the UK Schmidt
Telescope. The plates were processed into the present compressed
digital form with the permission of these institutions.}
%
% References
%

%

\begin{references}
\reference Blitz, L., Spergel, D.\ N., Teuben, P.\ J., Hartmann, D.,
 \& Burton, W.\ B.\ 1999, \apj, 514, 818
\reference Bureau, M., \& Carignan, C.\ 2002, \aj, 123, 1316
\reference Howard, S., \& Byrd, G.\ G.\ 1990, \aj, 99, 1798
\reference Karachentsev, I.\ D., et al.\ 2000, \aap, 363, 117
\reference Puche, D., Westpfahl, D., Brinks, E., \& Roy, J.-R.\ 1992,
 \aj, 103, 1841
\reference Putman, M.\ E., et al.\ 2002, \aj, 123, 873
\reference Quilis, V., Moore, B., \& Bower, R.\ 2000, Science, 288, 1617
\reference Rhode, K.\ L., Salzer, J.\ J., Westpfahl, D.\ J., \&
 Radice, L.\ A.\ 1999, \aj, 118, 323
\reference Rots, A.\ H., Crane, P.\ C., Bosma, A., Athanassoula, E.,
 \& van der Hulst, J.\ M.\ 1990, \aj, 100, 387
\reference Sargent, W.\ L.\ W., Sancisi, R., \& Lo, K.\ Y.\ 1983,
 \apj, 265, 711
\reference Schulz, S., \& Struck, C.\ 2001, \mnras, 328, 185
\end{references}
\end{document}